\documentclass[twocolumn,aps,amsmath,amssymb,prb]{revtex4}
\usepackage{graphicx}

\newcommand{\B}{\mbox{\tiny B}}
\newcommand{\ti}{\tilde}
\newcommand{\nl}{\nonumber \\}
\newcommand{\be}{\begin{equation}}
\newcommand{\ee}{\end{equation}}
\newcommand{\bea}{\begin{eqnarray}}
\newcommand{\eea}{\end{eqnarray}}
\newcommand{\bsube}{\begin{subequations}}
\newcommand{\esube}{\end{subequations}}
\newcommand{\Eq}[1]{Eq.\,(\ref{#1})}
\newcommand{\Eqs}[1]{Eqs.\,(\ref{#1})}
\newcommand{\Fig}[1]{Fig.\,\ref{#1}}

\newcommand{\dg}{\dagger}
\newcommand{\la}{\langle}
\newcommand{\ra}{\rangle}
\newcommand{\bfGam}{\mbox{\boldmath $\Gamma$}}

\begin{document}

\title{Exact quantum master equation via
   the calculus on path integrals}

\author{Rui-Xue Xu}
\author{Ping Cui}
\affiliation{
 Hefei National Laboratory for Physical Sciences at Microscale,
  University of Science and Technology of China, Hefei, China,
and \\
 Department of Chemistry, Hong Kong University of Science
  and Technology, Kowloon, Hong Kong
}
\author{Xin-Qi Li}
\affiliation{Institute of Semiconductors, Chinese Academy Sciences,
   P.O.\ Box 912,
  Beijing 100083, China, and \\
 Department of Chemistry, Hong Kong University of Science
  and Technology, Kowloon, Hong Kong
}
\author{Yan Mo}
\author{YiJing Yan}
\affiliation{Department of Chemistry, Hong Kong
  University of Science and Technology, Kowloon, Hong Kong}

\date{Submitted 1 September 2004; Revised 12 October 2004}
\begin{abstract}
   An exact quantum master equation formalism is constructed
  for the efficient evaluation of quantum non-Markovian dissipation
  beyond the weak system-bath interaction regime
  in the presence of time-dependent external field.
A novel truncation scheme is further proposed and compared with
other approaches to close the resulting hierarchically coupled
equations of motion. The interplay between system-bath interaction
strength, non-Markovian property, and required level of hierarchy
is also demonstrated with the aid of simple spin-boson systems.
\end{abstract}
\maketitle

   The key quantity in quantum statistical dynamics
is the reduced density operator, $\rho(t) \equiv {\rm
tr}_{\B}\rho_{\rm T}(t)$, i.e., the partial trace of the total
density operator over the bath space. It is well known that the
exact evolution of $\rho(t)$ can be formulated in terms of path
integral functional assuming the harmonic linear coupling bath
model. Its numerical implementation is however much tedious in
comparison with that of the differential quantum master equation
formulations.
 Tanimura {\it et al.}\ \cite{Tan89101,Tan914131} had
constructed exact quantum Fokker-Planck equations from path
integral formulations based on a Gaussian-Markovian bath model.
More recently, the construction of exact quantum master equations
has been achieved by exploiting various stochastic approaches to
dissipative
dynamics.\cite{Str991801,Yu9991,Str04052115,Yu04062107,Sha045053,Yan04216}

  In this article, we generalize the
Tanimura-Kubo's method\cite{Tan89101,Tan914131} to arbitrary cases
and construct an exact differential formulation of $\rho(t)$ based
on rather simple calculus on the path integral functional. Let us
start with the review of the path integral formulation. The total
Hamiltonian assumes $H_{\rm T}(t)=H(t)-\sum_aQ_aF_a(t)$. Here,
$H(t)$ is the deterministic part governing the coherent motion of
the reduced system, and the system-bath interaction is
characterized in terms of the coupling between system operators
$\{Q_a\}$ and stochastic bath  operators $\{F_a(t) \equiv
e^{ih_{\rm B}t}F_ae^{-ih_{\rm B}t}\}$. For an initial $\rho_{\rm
T}(t_0) = \rho(t_0)\rho^{\rm eq}_{\B}$; with $\rho_{\B}^{\rm eq}
\propto e^{-\beta h_{\rm B}}$ being the thermal bath density
operator, we have [$\hbar\equiv 1$ and $\beta \equiv 1/(k_BT)$
hereafter]
 \bea\label{rhot_def}
   \rho(t)
  &=& {\rm tr}_{\B} \Bigl[U_{\rm T}(t,t_0;\{F_{a}(t)\}) \rho(t_0)
      e^{-\beta h_{\rm B}}
    \nl && \qquad \times
        U^{\dg}_{\rm T}(t,t_0;\{F_{a'}(t)\})
             e^{\beta h_{\rm B}}\rho^{\rm eq}_{\B}\Bigr],
\eea where $U_{\rm T}(t,t_0;\{F_{a}(t)\})$ is the Hilbert-space
propagator with the total Hamiltonian $H_{\rm T}(t)$. Noting that
$e^{\beta h_{\rm B}}F_{a}(t)e^{-\beta h_{\rm B}}
  = F_{\rm a}(t-i\beta)$,
\Eq{rhot_def} can be recast as
\be \label{rhot_def1}
  \rho(t) =
    \la U_{\rm T}(t,t_0;\{F_{a}(t-i\beta)\})
     \rho(t_0) U^{\dg}_{\rm T}(t,t_0;\{F_{a'}(t)\}) \ra.
\ee
 Here, $\la\,\cdot\,\ra\equiv{\rm tr}_{\B}(\,\cdot\,\rho^{\rm eq}_{\B})$.
Denote
 \be \label{Udef}
  \rho(t) \equiv {\cal U}(t,t_0) \rho(t_0) ,
 \ee
which defines the dissipative Liouville-space propagator ${\cal U}(t,t_0)$.
 Let $\bfGam\equiv\{\alpha,\alpha'\}$ be an arbitrary
Liouville-space representation for the reduced system.
The path integral formulation for
 ${\cal U}(\bfGam_t,t;\bfGam_0,t_0) \equiv
  \la\la \bfGam_t|{\cal U}(t,t_0)|\bfGam_0\ra\ra$
is then\cite{Fey63118}
 \be \label{dissU}
   {\cal U}(\bfGam_t,t;\bfGam_0,t_0)
    = \int_{\bfGam_0}^{\bfGam_t}\!\!\!{\cal D}\bfGam
   e^{iS[\alpha]}{\cal F}[\alpha,\alpha']e^{-iS[\alpha']}.
 \ee
Here, $S[\alpha]$ and  ${\cal F}[\alpha,\alpha']$
denote the action and influence functionals, respectively.
The latter is given as [cf.~\Eq{rhot_def1}]
 \bea\label{FV_fun0}
  {\cal F}[\alpha,\alpha']=
 \Bigl\la
   \exp_{+}\bigl\{  i \sum_{a}
      \int_{t_0}^t\!d\tau\, Q_a[\alpha(\tau)] F_{a}(\tau-i\beta)  \bigr\}
\nl\times
   \exp_{-}\bigl\{ -i \sum_{a'}
      \int_{t_0}^t\!d\tau\, Q_{a'}[\alpha'(\tau)] F_{a'}(\tau)  \bigr\}
 \Bigr\ra.
 \eea
For the stochastic bath operators $\{F_a(t)\}$ that satisfy the
Gaussian statistics, the above equation can be recast in terms of
the standard Feynman-Vernon form:\cite{Fey63118}
 \bea\label{FV_fun}
   {\cal F}[\alpha,\alpha'] &=&
  \exp\Bigl(-
   \sum_a
    \int_{t_0}^{t}\!d\tau
      \bigl\{ Q_a[\alpha(\tau)]-Q_a[\alpha'(\tau)] \bigr\}
    \nl && \qquad \times
      \bigl\{ \ti Q_a[\alpha(\tau)] -\ti Q^{\ast}_a[\alpha'(\tau)] \bigr\}
  \Bigr),
\eea
with
\be \label{tilQ}
  \tilde Q_a[\alpha(t)] = \sum_{b} \int_{t_0}^{t}\!d\tau\,
    C_{ab}(t-\tau) Q_b[\alpha(\tau)].
\ee
 Here, $C_{ab}(t-\tau) \equiv \la F_{a}(t)  F_{b}(\tau) \ra$
are the bath correlation functions, which satisfy the symmetry and
detailed-balance relations $C^{\ast}_{ab}(t) = C_{ba}(-t) =
C_{ab}(t-i\beta)$.

   We are now in the position to derive the
exact quantum master equation via a rather simple calculus on the
above integral functional formulation. For clarity, we shall in
the following consider only the single dissipative mode case, in
which the system-bath interaction assumes $H'(t)=-QF(t)$. Consider
the time derivative of \Eq{dissU} term by term. Firstly, \be
\label{leftH}
  \frac{\partial}{\partial t} e^{iS[\alpha]} = -i\!\int\!d\bar\alpha
    H_t(\alpha,\bar\alpha) e^{iS[\bar\alpha]} \equiv
     -iH(t)\!\cdot\!e^{iS[\alpha]}.
\ee
The last identity defines the notion used in this paper.
We have also
\be \label{rightH}
 \frac{\partial}{\partial t} e^{-iS[\alpha']}
  = ie^{-iS[\alpha']}\!\cdot\!H(t).
\ee
Similarly,
\bea \label{dotF}
  \frac{\partial}{\partial t} {\cal F}
=
  - Q \!\cdot\![(\ti Q_t - \ti Q_t^{' \ast}){\cal F}]
  + [(\ti Q_t - \ti Q^{' \ast}_t){\cal F}]\!\cdot\!Q ,
\eea
with
\be \label{tiQt}
   \ti Q_t \equiv \int_{t_0}^t\!d\tau\, C(t-\tau)Q[\alpha(\tau)].
\ee
  Together with \Eqs{Udef} and (\ref{dissU}), we obtain
\be \label{qme0}
  \dot\rho(t) = -i[H(t),\rho(t)] -i [Q,\rho_1(t)].
\ee
Here, $\rho_1(t) \equiv {\cal U}_1(t,t_0)\rho(t_0)$,
or generally,
\be \label{rhon}
   \rho_n(t) \equiv {\cal U}_n(t,t_0)\rho(t_0),
\ee
with ${\cal U}_n(t,t_0)$ being given in terms of
path integral as
\be \label{Un}
  {\cal U}_n(\bfGam_t,t;\bfGam_0,t_0) =
      \int_{\bfGam_0}^{\bfGam_t}\!\!\!{\cal D}\bfGam
   e^{iS[\alpha]}{\cal F}_n[\alpha,\alpha']e^{-iS[\alpha']}   ,
\ee
and
\be \label{Fn}
 {\cal F}_n[\alpha,\alpha']
\equiv
  (-i)^n (\ti Q_t - \ti Q^{' \ast}_t)^n {\cal F}[\alpha,\alpha'].
\ee
Note that $\rho_0(t) = \rho(t)$.
We have further [cf.\ \Eq{dotF}]
\bea \label{dotFn}
   &\ & \frac{\partial}{\partial t}
     {\cal F}_n[\alpha,\alpha']
\nl &=&
   -i \left\{Q\!\cdot\!{\cal F}_{n+1}[\alpha,\alpha']
   - {\cal F}_{n+1}[\alpha,\alpha'] \!\cdot\! Q \right\}
\nl &\ &
   -i n\left\{C(0)Q\!\cdot\!{\cal F}_{n-1}[\alpha,\alpha']
   - C^{\ast}(0){\cal F}_{n-1}[\alpha,\alpha']\!\cdot\!Q\right\}
\nl &\ &
   + n \ti{\cal F}_{n}[\alpha,\alpha'] ,
\eea
where
\bea \label{tiFn}
  \ti{\cal F}_{n}[\alpha,\alpha']
&\equiv&
 -i  {\cal F}_{n-1}[\alpha,\alpha']
   \int_{t_0}^{t}\!d\tau
    \big\{\dot C(t-\tau)Q[\alpha(\tau)]
\nl &&  \qquad \qquad
    - \dot C^{\ast}(t-\tau)Q[\alpha'(\tau)]\big\}  .
\eea
 We have therefore for $\rho_n; n\geq 0$ [cf.~\Eqs{rhon} and (\ref{Un})]
 \bea\label{dotrhon}
  \dot\rho_n(t)
&=&
 -i[H(t),\rho_n(t)]-i [Q,\rho_{n+1}(t)]
\nl & &
  -in \left[C(0)Q\rho_{n-1}(t)
  -C^{\ast}(0)\rho_{n-1}(t)Q\right]
\nl & &
  + n \ti\rho_{n}(t) .
\eea
Here,
\be \label{tirho}
  \ti\rho_{n}(t) \equiv \ti{\cal U}_n(t,t_0)\rho(t_0) ,
\ee
with $\ti{\cal U}_n(t,t_0)$ being given in terms of
path integral as
\be \label{tiUn}
  \ti{\cal U}_n(\bfGam_t,t;\bfGam_0,t_0) =
      \int_{\bfGam_0}^{\bfGam_t}\!\!\!{\cal D}\bfGam
   e^{iS[\alpha]}\ti{\cal F}_n[\alpha,\alpha']e^{-iS[\alpha']}   .
\ee
Both $\rho_n$ [\Eq{rhon} with \Eqs{Un} and (\ref{Fn})]
and $\ti\rho_n$ [\Eq{tirho} with \Eqs{tiUn} and (\ref{tiFn})] are of
the $(2n)^{\rm th}$--order in the system-bath interaction as their
leading contributions. In \Eq{dotrhon}, $\rho_n(t)$
depends on $\rho_{n\pm 1}(t)$  and $\ti\rho_n(t)$.
The latter however does not belong to the same hierarchy.
To proceed, we shall first convert \Eqs{dotrhon} into a
set of hierarchically coupled equations of motion (EOM),
followed by a proper truncation scheme to
close the resulting EOM.
 Note that in some special cases, such as the driven
 Brownian oscillators\cite{Xu037,Hu922843,Kar97153} and
 the pure-dephasing (non-demolishing) dynamics,
 the exact $\rho_1(t)$ can be
 obtained in terms of $\rho_0(t) =\rho(t)$; thus the
 non-hierarchical exact quantum master equation can be constructed.

  The hierarchicalization of  \Eqs{dotrhon}
can be carried out with certain forms of bath correlation
function. Consider, for example, a Gaussian-Markovian bath with
$C(t)=\eta e^{-\gamma t}$, where $\gamma$ is real and $\eta$ may
be complex. In this case, $\ti\rho_n(t) = -\gamma\rho_n(t)$ and
\Eqs{dotrhon} become completely hierarchical, recovering the
Fokker-Planck equation obtained previously by Tanimura and
co-workers.\cite{Tan89101,Tan914131}

 Consider another model in which the bath
spectral density assumes  $J(\omega)\propto
\omega/(\omega^2+\gamma^2)^2$, and the resulting bath correlation
is found to be of $C(t)=\nu_0te^{-\gamma t}+\nu_1e^{-\gamma t}$,
if the Matsubara terms can be neglected.\cite{Xu037,Yan05} Here,
$\gamma$ is real and $\nu_0$ and $\nu_1$ are complex. The
hierarchically coupled EOM equivalent to \Eq{dotrhon} can readily
be constructed as follows. For the present model of $C(t)$, we
have $\ti Q_t=\nu_0\ti Q_0+\nu_1\ti Q_1$ [cf.~\Eq{tiQt}], where
$\ti Q_0 = \int_{t_0}^t\!d\tau\,
      (t-\tau)e^{-\gamma(t-\tau)}Q[\alpha(\tau)]$ and
$\ti Q_1 = \int_{t_0}^t\!d\tau\,
      e^{-\gamma(t-\tau)}Q[\alpha(\tau)]$.
Consequently, ${\cal F}_n$ of \Eq{Fn} can be
expressed in terms of
${\cal F}^{k'j'}_{kj} \equiv
{\ti Q}_0^k  (\ti Q'_0)^{k'} {\ti Q}_1^j  (\ti Q'_1)^{j'} {\cal F}$,
with $k+j+k'+j'=n$.
Introduce now $\rho^{k'j'}_{kj}$ similarly as
 \Eqs{rhon} and (\ref{Un}) but with ${\cal F}_n$ there being replaced
by ${\cal F}^{k'j'}_{kj}$.
Note that $\partial\ti Q_0/\partial t =\ti Q_1-\gamma\ti Q_0$ and
      $\partial\ti Q_1/\partial t=Q_t-\gamma\ti Q_1$.
The hierarchical EOM for
$\rho^{k'j'}_{kj}$;  with  $\rho_{00}^{00}\equiv \rho$,
can then be obtained as
\bea\label{dotrhonHK}
 \dot \rho^{k'j'}_{kj}
&=& -i[H(t),\rho^{k'j'}_{kj}]-(k+j+k'+j')\gamma\rho^{k'j'}_{kj}
\nl &&
  -\bigl[Q,\nu_0\rho^{k'j'}_{k+1,j}-\nu_0^\ast\rho^{k'+1,j'}_{kj}\bigr]
\nl &&
  -\bigl[Q,\nu_1\rho^{k'j'}_{k,j+1}-\nu_1^\ast\rho^{k',j'+1}_{kj}\bigr]
\nl &&
  +k\rho^{k'j'}_{k-1,j+1}+k'\rho^{k'-1,j'+1}_{kj}
\nl &&
  +jQ\rho^{k'j'}_{k,j-1}+j'\rho^{k',j'-1}_{kj}Q .
\eea Obviously, $\rho^{k'j'}_{kj}$ is of the $[2(k+j+k'+j')]^{\rm
th}$--order in the system-bath interaction as its leading
contribution. In general, the bath correlation function $C(t)$ can
be expressed in terms of a series of exponential expansion, and
thus, by following the similar procedure as shown above, one can
construct a set of hierarchically coupled EOM based on
\Eq{dotrhon}.

  To complete the formulation, we propose in the following
  a novel truncation scheme and compare it with
  other two existing approaches to close the hierarchical EOM
  equivalent to \Eq{dotrhon}.
 As shown earlier $\ti\rho_n(t)$ in \Eq{dotrhon}
 is of the same order as $\rho_n(t)$, we need only to
 approximately anchor $\rho_{N}$ in terms of $\rho_{n < N}$,
 such that the  system-bath interactions
 are accounted rigorously for up to $(2N)^{\rm th}$
 order, but partially for higher-order
 contributions.
  To do that, let us first recast \Eq{Fn} as
${\cal F}_{n}=-i(\ti Q_t - \ti Q^{'\ast}_t){\cal F}_{n-1}$, and
then approximate the involving $\ti Q_t $ with its
bath-free-evolution counterpart; i.e.,
 \be\label{tiQquant}
   {\ti Q}_{\rm trun}(t)\equiv\int_{t_0}^t\!d\tau\,
      C(t-\tau) {\cal G}(t,\tau) Q.
 \ee
 Here, ${\cal G}(t,\tau)$ denotes the dissipation-free
propagator that satisfies $\partial{\cal G}(t,\tau)/\partial t
=-i{\cal L}(t){\cal G}(t,\tau)$, with the reduced system
Liouvillian ${\cal L}(t)\cdot \equiv [H(t),\cdot]$. The resulting
truncation scheme proposed here amounts therefore to
[cf.~\Eqs{rhon} and (\ref{Un})]
\be\label{rhontrunc}
  \rho_{N}(t)
\approx
  -i\left[{\ti Q}_{\rm trun}(t)\rho_{N-1}(t)
  -\rho_{N-1}(t){\ti Q}_{\rm trun}^\dg(t)\right].
\ee
 The above truncation leads to a time-local evolution for
 the anchoring $\rho_{N}$, and the resulting EOM is in a
 partial ordering prescription (POP).

  In comparison, the corresponding chronological ordering
  prescription (COP) of $(2N)^{\rm th}$-order truncation can
  be realized via setting $\rho_{N+1} = 0 $, as it effectively leads
   to a time-ordered, memory description
   of $\rho_{N}(t)$ in terms of $\rho_{n<N}(\tau\leq t)$.
  It is easy to verify that the conventional second-order
  memory-kernel and time-local formulations\cite{Mei993365,Muk81509,Yan982721,Xu029196}
  are just the special cases of the present
   COP-truncation ($\rho_2=0$) and POP-truncation [\Eq{rhontrunc} with $N=1$],
  respectively.

  Consider now the truncation scheme
  proposed by Tanimura and co-workers,\cite{Tan89101,Tan914131}
  in which the fast modulation or Markovian anzatz
  is assumed applicable at the anchor level.
  In this case,
  $\ti Q_{\rm trun}(t)$ of \Eq{tiQquant} reduces to
  $\ti Q^{\rm mar}_{\rm trun}={\bar C} Q$,
  where $\bar C=\int_0^\infty \!\! d\tau C(\tau)$.
  This scheme is essentially identical
  to that proposed by Tanimura and co-workers, but
  is obtained here for general $C(t)$ without
  invoking the Gaussian-Markovian bath model.
  Unlike the $N$-level
  POP  [\Eq{rhontrunc} with
  \Eq{tiQquant}] or COP ($\rho_{N+1}=0$) truncation,
  the $N$-level fast-modulation truncation scheme
  is not a rigorous $(2N)^{\rm th}$, but rather
  $(2N-2)^{\rm th}$-order
  formulation. We shall come back to this point later
  in terms of  the practical efficiencies of
  the aforementioned three truncation schemes; see discussions following \Fig{fig1}.
  Obviously, all these schemes become exact
  when $N \rightarrow \infty$.

    To investigate the efficiencies of involving truncation schemes
  for the study of non-Markovian dissipation with
  arbitrary system-bath interaction,
  we consider here a simple spin-boson system: $H=\frac{1}{2}\Omega\sigma_x$,
  with $Q=\sigma_z$ and $C(t)=\Delta^2 e^{-\gamma t}$.
  In the following, we adopt the integrated coupling
  strength $\Gamma\equiv \int_0^{\infty}\!dt C(t)=\Delta^2/\gamma$
  and the dimensionless modulation parameter
  $\kappa\equiv \gamma/\Delta$ to characterize
  the nature of system-bath coupling,
  and $s_N \equiv \frac{1}{T} \max_{\{t_i\}}\Bigl\{\int_{t_i}^{t_i+T}\!
   dt|A_N(t)-A_{N\rightarrow\infty}(t)|^2\Bigr\}$; with
   $T\equiv 2\pi/\Omega$, to calibrate the finite $N$-truncation induced
   error in evaluation of a given test quantity $A(t)$, which
   will be chosen to be $\rho_{22}(t)-\rho_{11}(t)$ for demonstration.
   The initial condition is set to be
   $\rho_{jk}(0)=\delta_{2j}\delta_{jk}$; $j,k=1,2$.

   Depicted in \Fig{fig1} are the resulting
   error indicators $s_N$ for the
   aforementioned three truncation schemes, demonstrated as
   functions of anchoring index $N$
   at various values of system-bath coupling strength
   and modulation parameter:
  (a) $\Gamma=\Omega$, $\kappa=1$;
  (b) $\Gamma=\Omega$, $\kappa=0.1$;
  (c) $\Gamma=10\,\Omega$, $\kappa=1$; and
  (d) $\Gamma=10\,\Omega$, $\kappa=0.1$.
  This figure clearly demonstrates the following features: (i)
  The anchor index $N$ depends not only on the
  system-bath coupling strength $\Gamma$, but more
  importantly on the modulation parameter $\kappa$;
  (ii) In the slow modulation ($\kappa\ll 1$) limit, the low order
  truncation may not be sufficient even in the weak
  coupling ($\Gamma \ll \Omega$) regime;
  (iii)
  As inferred from their constructions, these three truncation
  schemes are in principle of the same quality in the
  fast modulation limit; see \Fig{fig1}(c).
  Considering further the computation efforts involved,
  while the anchor $\rho_N$ in the POP or fast-modulation
  truncation is expressed directly by \Eq{rhontrunc} or its
  Markovian counterpart, respectively, the $\rho_N^{\rm COP}$
  shall however be propagated via coupled EOM.
  Thus, the POP
  truncation scheme proposed in the work is overall the best,
  including the case studied in \Fig{fig1}(b).

  Shown in \Fig{fig2} are the approximate results of
  the test quantity $\rho_{22}(t)-\rho_{11}(t)$,
  obtained via the POP (solid),  the COP (dash),
  and the fast-modulation (dot) truncation schemes
  at the specified value of $N$ in each panel
  such that the error tolerance of $10^{-3.5}$ is met by the
  minimum $s_N$ among these three truncation
  schemes. Included in each panel of this figure is also
  the exact result (thick-solid) that can be obtained
  via any truncation scheme with a sufficiently large $N$.
  Note that \Fig{fig2}(b) is depicted for  $26<t<27$ (in the unit
  of $2\pi/\Omega$) where the difference between various curves is relatively large.
  The overall superiority of POP-truncation scheme is
  again highlighted. We have also showed (iv) The behavior of $\rho(t)$
  in short time, long time, and Markovian regimes may be accounted
  for properly by the low order $\rho_n$;
  (v) The relatively high order $\rho_n$ are required for the
  intermediate time, and the slow modulation regimes;
  (vi) Note that
   the critical damp occurs at $\Gamma=\Omega$ in
   the fast modulation ($\kappa\gg 1$) limit
   [cf.\ \Fig{fig2}(a)], it occurs
   however at a much larger $\Gamma$
   in the slow modulation ($\kappa\ll 1$) limit
   [cf.\ \Fig{fig2}(b)].

 To summarize, we have derived an exact quantum master equation
 formalism via a
 simple calculus on the Feynman-Vernon influence functional path
 integrals. It is valid for
  arbitrary external field driven reduced dynamics
  under non-Markovian system-bath interaction
  beyond the weak coupling regime.
  The present derivation constitutes an alternative
  approach to the exact quantum master equation, \Eq{dotrhon},
  which is identical to that obtained recently by Shao
  via rather advanced stochastic differential equation
   algebra.\cite{Sha045053,Yan04216,Oks98}
   It is also noticed that Shi and Geva had also recently
   constructed a formally exact quantum dissipation
   theory.\cite{Shi0312063}
   It however depends practically on the path-integral
   evaluation of the Nakajima-Zwanzig dissipation kernel,
   despite the fact that is expressed in terms of force-force
   correlation function to be evaluated with correlated
   system-bath ensemble.\cite{Shi0312063}

  We have also proposed a novel truncation scheme; i.e.,
  the POP-scheme of \Eq{rhontrunc}, which is shown to be
  overall superior to the two existing truncation approaches.
  This superiority is of much more implication than that
  demonstrated in the numerical examples of this work, as
  in general $C(t)$ contains multi-exponential terms,
  especially at low temperature regime, and the relevant number
  of $\rho_{n}$; with $n\leq N$ [cf.\ \Eq{dotrhonHK}]
  in the hierarchical EOM increases quasi-exponentially with the
  truncation anchor $N$. The numerical demonstrations are also of
  rich physical implications on the interplay between system-bath
  coupling strength, non-Markovian property, and the required
  order of truncation, which had been summarized
  in details in Comments (i)--(vi)
  following \Fig{fig1} and \Fig{fig2}.
  Especially, low-order quantum dissipation theories
  should be used with care if non-Markovian dynamics is
  important.

    Support from the RGC Hong Kong and
    the CAS Foundation for Outstanding Overseas Scholars (Y.J.Yan),
    the NNSF of China No.\ 20403016 (R.X.Xu),
    and the Major State Basic Research Project No.\ G001CB3095 of China
    and the CAS Special Fund
    for ``100 Person'' Project (X.Q.Li) is acknowledged.
    The authors also thank Professor Jiushu Shao
    for great inspiration and critical comments.


\begin{figure}
\caption{The error parameters $s_N$ versus the anchor $N$, in the
evaluations of $\rho_{22}(t)-\rho_{11}(t)$ via the POP, COP, and
fast-modulation truncation schemes. The selected system-bath
coupling strength
   and modulation parameters are:
  (a) $\Gamma=\Omega$, $\kappa=1$;
  (b) $\Gamma=\Omega$, $\kappa=0.1$;
  (c) $\Gamma=10\,\Omega$, $\kappa=1$; and
  (d) $\Gamma=10\,\Omega$, $\kappa=0.1$.}
\label{fig1}
\end{figure}

\begin{figure}
\caption{The evolutions of $\rho_{22}(t)-\rho_{11}(t)$ implied in
\Fig{fig1}, with the specified anchor $N$ in each panel being
chosen such that the tolerance of $10^{-3.5}$ is satisfied by the
minimum $s_N$ among the three truncation schemes. The exact
evolutions can be obtained with any schemes truncated at
sufficiently high level. The curves for the exact and the POP
almost overlap with each other in each of the four panels.}
 \label{fig2}
\end{figure}

\clearpage
\includegraphics{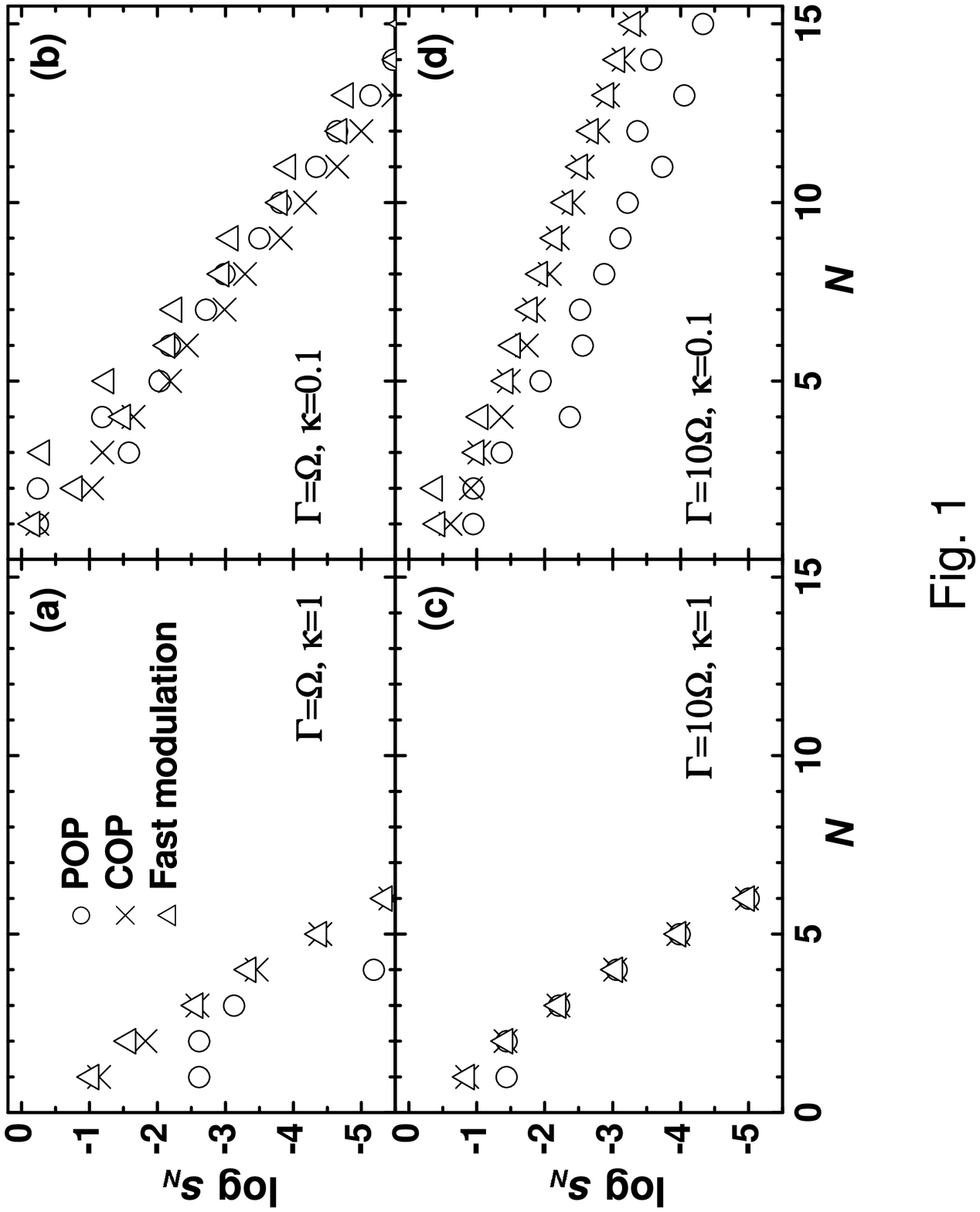}
\clearpage
\includegraphics{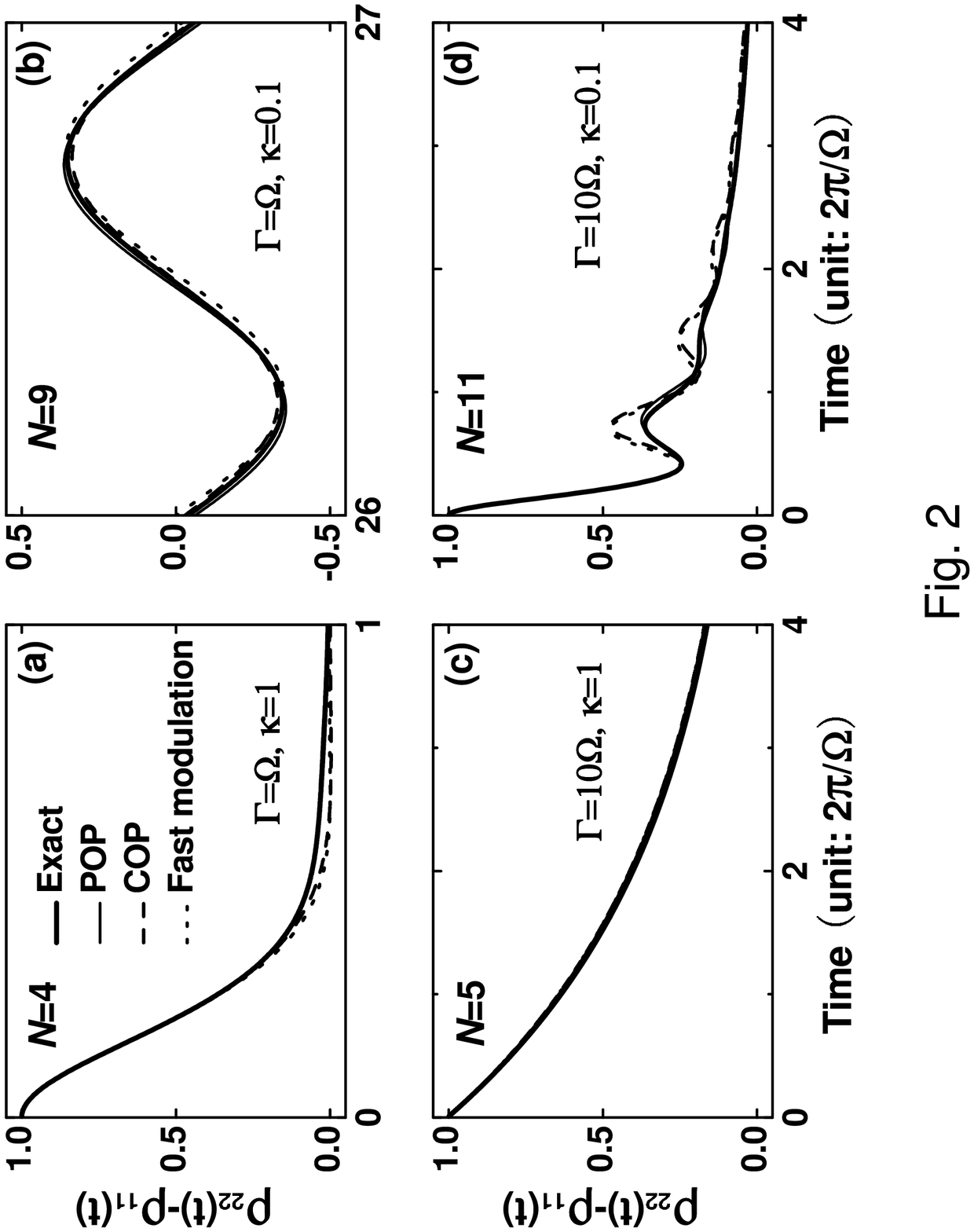}

\end{document}